\documentclass[12pt]{article}
\usepackage{amsmath}
\usepackage{latexsym}
\usepackage{epsf}
\usepackage{amsfonts}
\author{Yungui Gong\\
Physics Department, The University of Texas at Austin, Austin, 
Texas 78712\\}
\title{On Holography in Brans-Dicke Cosmology\footnote{Talk given at the 9th Midwest Relativity
Conference held in University of Illinois at Urbana Champaign,
November 12-13, 1999.}}

\begin{document}

\maketitle
\begin{flushright}
UTEXAS-HEP-99-20
\end{flushright}
\parindent=4ex
\begin{abstract}
The holographic bound in Brans-Dicke $k=1$ matter dominated Cosmology is discussed. 
In this talk, both the apparent
horizon and the particle horizon are taken for the holographic bound. 
The covariant entropy conjecture proposed by Bousso is also 
discussed. 
\end{abstract}
\section{Introduction}
Dilaton field appears naturally both from 
the Kaluza-Klein compactification
and the String spectrum.
The simplest way to incorporate the scalar field as the spin-0 
partner of spin-2 gravitational
field is Brans-Dicke theory in which the gravitational coupling
constant is replaced by a scalar field. In \cite{gong}, the author
considered the holographic bound in the region within the particle
horizon for a general Brans-Dicke universe. The discussion for the
case $k=1$ matter
dominated is not complete because there is no analytical solution
for that case. In this paper, we will investigate the holographic bound
for the $k=1$ matter dominated Brans-Dicke universe in regions within particle
horizon and apparent horizon. The particle horizon idea was first 
proposed by Fischler and Susskind \cite{fs}. 
Because the holographic bound in the region
within the particle horizon is not satisfied for closed matter dominated universe,
Bak and Rey used the apparent horizon to solve the problem \cite{dbsr}. 
However, Kaloper and Linde showed that the holographic bound
in the regions within apparent horizon could be violated in the
anti-De-Sitter space \cite{nkal}. Bousso's covariant entropy conjecture gave a means
to select a null hypersurface starting from any 2 dimensional spacelike
surface so that the holographic bound would be satisfied in the null
hypersurface \cite{bousso}. This holographic bound can be applied to general spacetime
starting from any surface.
A generalized version of this conjecture was proved by Flanagan, Marolf
and Wald under some assumptions on the entropy flux \cite{fmw}.
Other cosmological
entropy bounds were discussed in \cite{cosmo}.

The Friedman-Roberson-Walker metric for $k=1$ is
\begin{equation}
\label{metric}
\begin{split}
ds^2
&=-dt^2+a^2(t)\left({dr^2\over 1-r^2}+r^2 d\Omega^2\right)\\
&=-dt^2+a^2(t)(d\chi^2+\sin^2\chi\, d\Omega^2),
\end{split}
\end{equation}
where $r=\sin \chi$. Here I use $c=\hbar=1$.
The standard cosmological equation and solutions are 
\begin{gather}
H^2 + {1\over a^2}={8\pi G\over 3}\rho,\\
\rho a^3=\rho_0 a^3_0,\\
a(\eta)={a_{max}\over 2}(1-\cos\eta)=a_{max}\sin^2(\eta/2),
\end{gather}
where the cosmic time $\eta$ is $d\eta=dt/a(t)$ 
and $a_{max}=8\pi G\rho_0 a^3_0/3$.
Note that $\eta\rightarrow \pi$, $a\rightarrow a_{max}$.
The particle horizon is
\begin{equation}
\label{parthor}
\chi_{PH}=\int_0^t {d{\tilde t}\over a({\tilde t})}=\eta.
\end{equation}
The apparent horizon is
\begin{gather}
r_{AH}={1\over a(t)\sqrt{H^2+1/a^2(t)}}=|\sin(\eta/2)|,\\
\label{apphor}
\chi_{AH}={\eta\over 2}.
\end{gather}

The idea of holographic bound is that the matter entropy 
inside a spatial region $V$
does not exceed 1/4 of the area $A$
of the boundary of that region measured in Planck units.
From the metric (\ref{metric}), the holographic bound in a region
with radius $r=\sin\chi$ for closed
universe is
\begin{equation}
\label{bound}
{S\over GA/4}={\epsilon V\over GA/4}={\epsilon(2\chi-\sin 2\chi)\over
G a^2(t)\sin^2\chi} \le 1,
\end{equation}
where $\epsilon$ is the constant comoving entropy density.

If we consider the spherical region inside the particle horizon
$\chi_{PH}=\eta$,
then the holographic bound (\ref{bound}) becomes
$${S\over GA/4}={\epsilon(2\eta-\sin 2\eta)\over
G a_{max}^2\sin^4(\eta/2)\sin^2\eta} \le 1.$$
It is obvious that the bound is violated when $\eta\rightarrow \pi$.
Therefore, the holographic bound proposed by Fischler and Susskind
does not apply to the closed universe.

If we consider the spherical region inside the apparent horizon
$\chi_{AH}=\eta/2$, the holographic bound (\ref{bound}) becomes
$${S\over GA/4}={\epsilon(\eta-\sin \eta)\over
G a_{max}^2\sin^6(\eta/2)} \le 1.$$
Therefore, the holographic bound is satisfied if it
is satisfied initially. So the holographic bound proposed
by Bak and Rey applies to the closed universe.

\section{Brans-Dicke Cosmology}
The Brans-Dicke Lagrangian in the Jordan frame is given by
\begin{equation}
\label{bdlagr}
{\cal L}_{BD}={\sqrt{-\gamma}\over 16\pi}\left[\phi{\tilde  R}-
\omega\,\gamma^{\mu\nu}
{\partial_\mu\phi\partial_\nu\phi\over \phi}\right]-{\cal L}_m(\psi,\,
\gamma_{\mu\nu}),
\end{equation}
with $\langle\phi\rangle=1/G$.

The cosmological equations are
\begin{gather}
\label{jbd1}
H^2+{k\over a^2}+H{\dot{\phi}\over \phi}-{\omega\over 6}\left({\dot{\phi}
\over \phi}\right)^2={8\pi\over 3\phi}\rho,\\
\label{jbd2}
\ddot{\phi}+3H\dot{\phi}={8\pi\over 2\omega +3}(\rho-3p),\\
\label{jbd3}
\rho a^3=\rho_0 a^3_0.
\end{gather}

\begin{figure}[htb]
\vspace{-0.1in}
\begin{center}
$\begin{array}{cc}
\epsfxsize=2.7in \epsffile{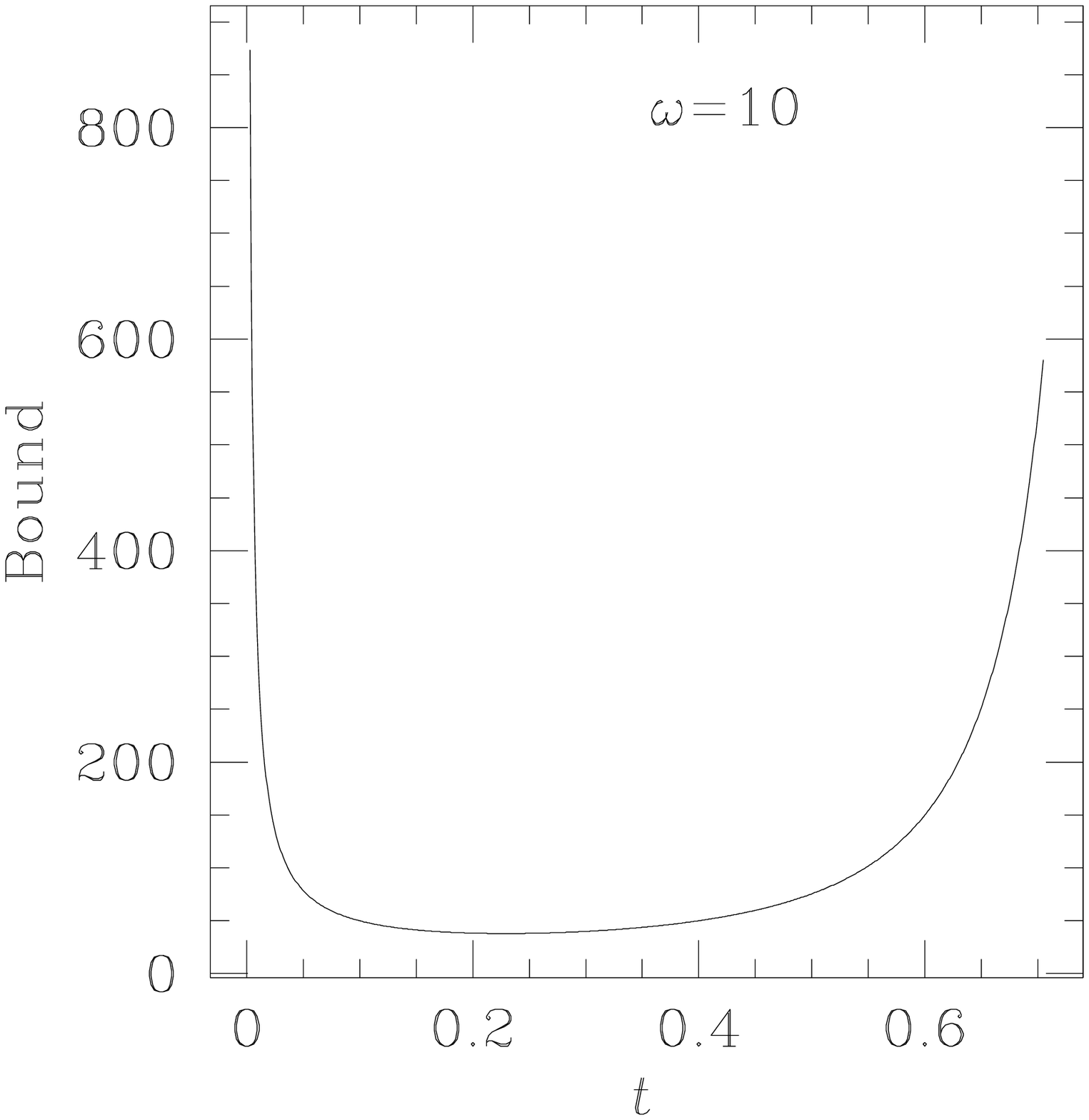}&
\epsfxsize=2.7in \epsffile{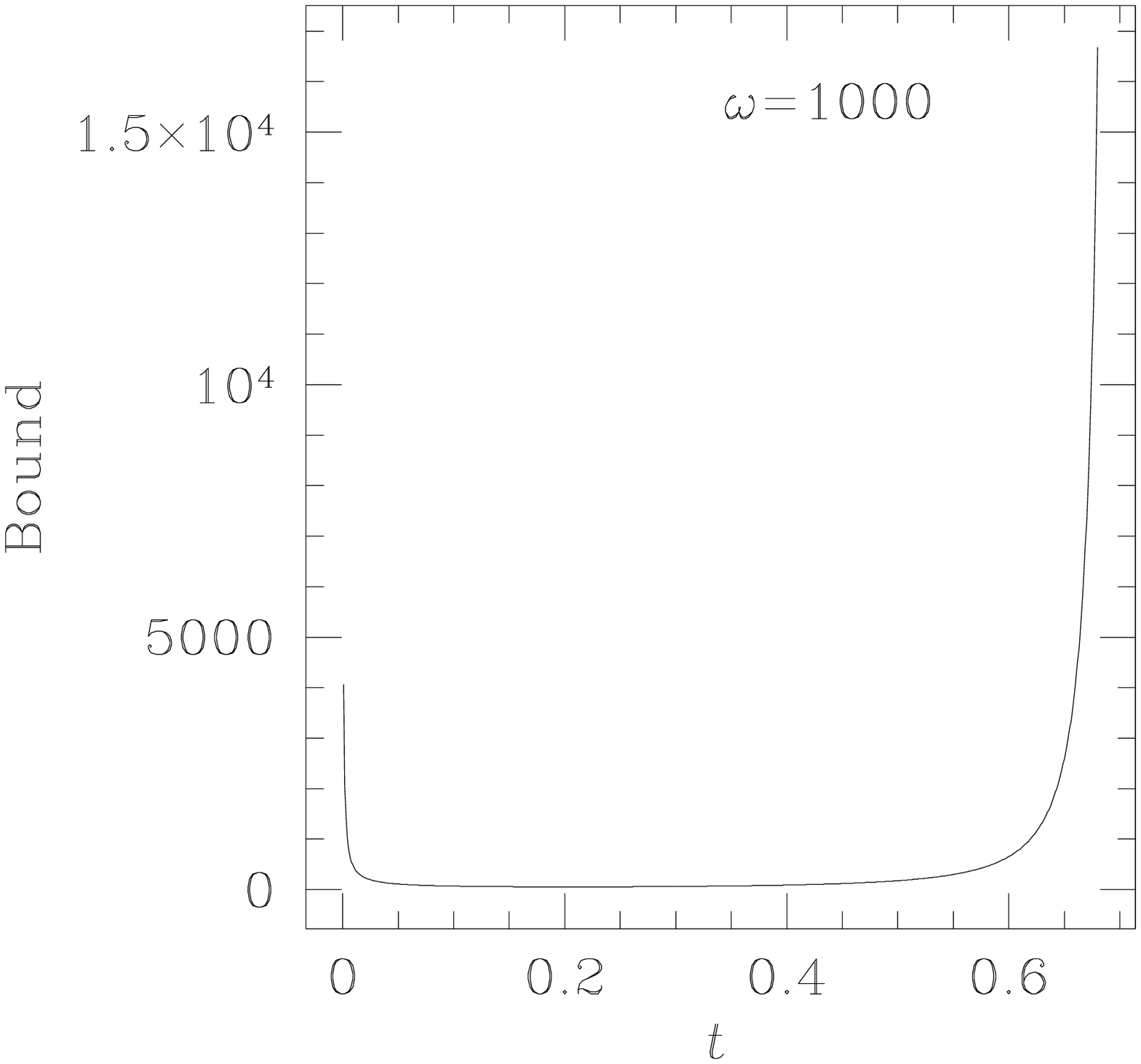}
\end{array}$
\end{center}
\vspace{-0.5in}
\caption{The holographic bound for particle horizon in Jordan frame with $\omega=10$ and 
$\omega=1000$ respectively.}
\label{phjordan}
\end{figure}

\begin{figure}[htb]
\vspace{-0.1in}
\begin{center}
$\begin{array}{cc}
\epsfxsize=2.7in \epsffile{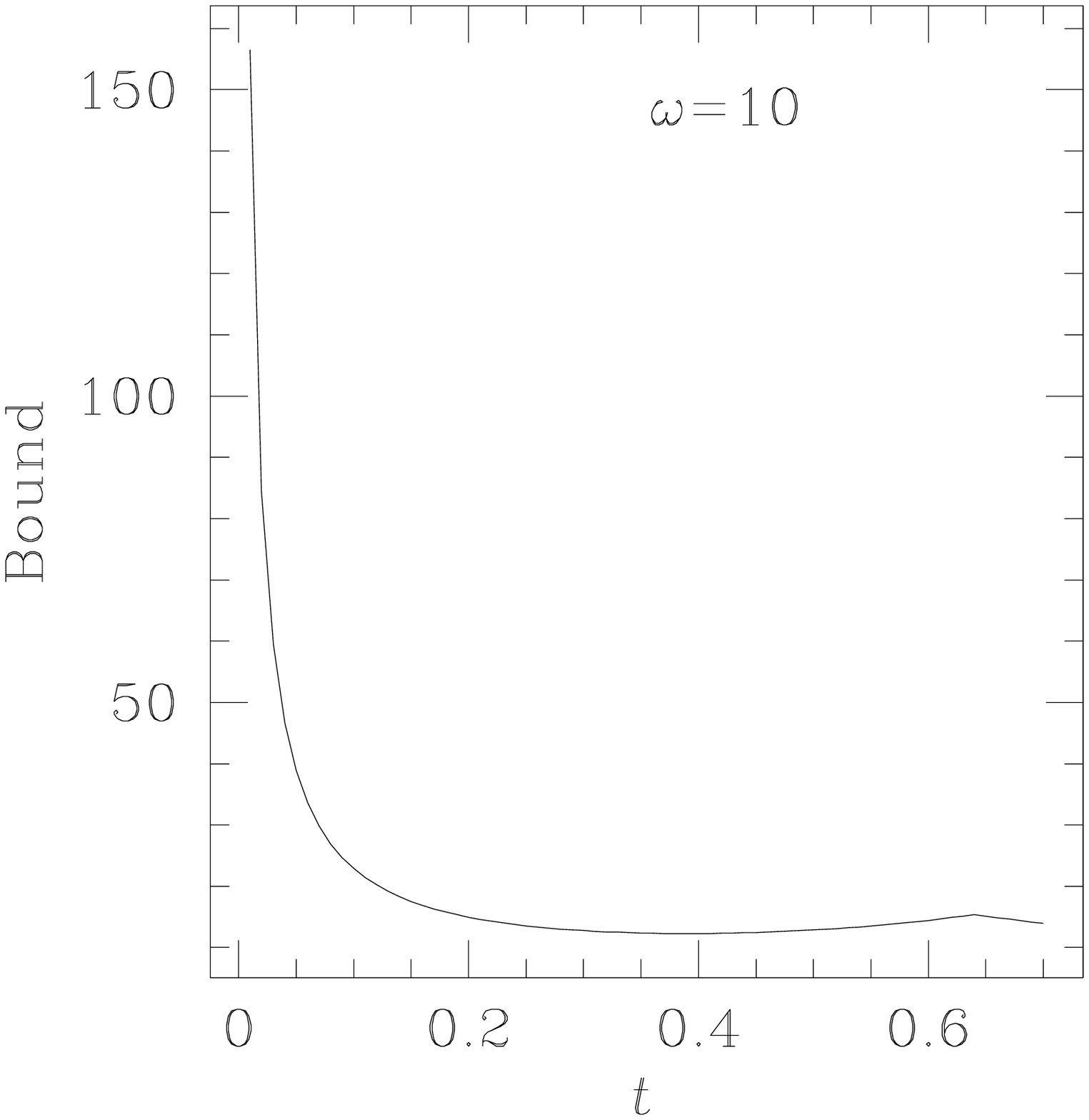}&
\epsfxsize=2.7in \epsffile{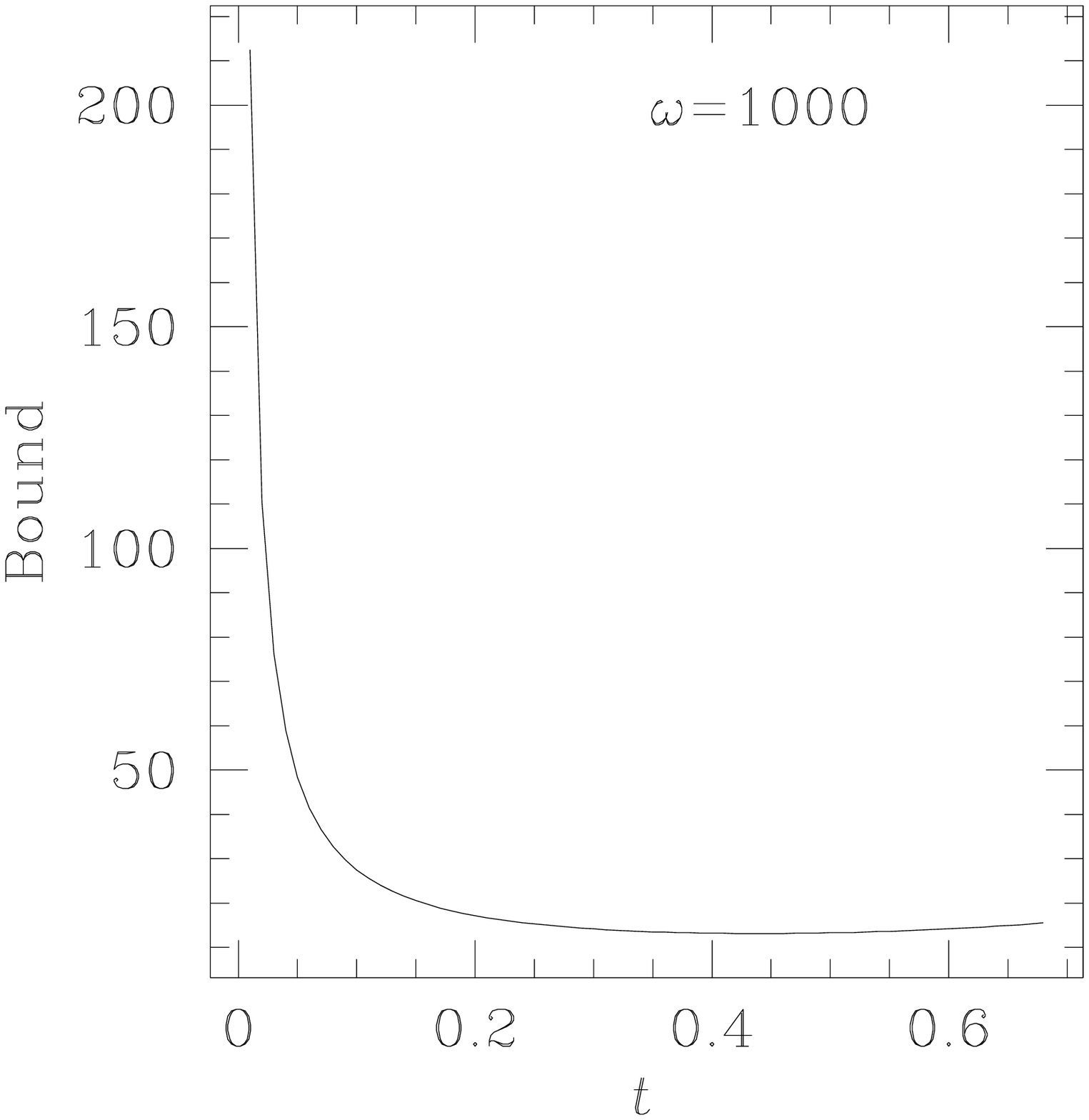}
\end{array}$
\end{center}
\vspace{-0.5in}
\caption{The holographic bound for apparent horizon in Jordan frame with $\omega=10$ and 
$\omega=1000$ respectively.}
\label{ahjordan}
\end{figure}

To solve the above equations for the case $k=1$ and $p=0$,
we take the solutions
for $k=0$ as the initial conditions.
\begin{gather}
\label{jbd5}
a(t_i)=t_i^p,\quad \phi(t_i)={4\pi(4+3\omega)\over 2\omega +3}\,t_i^q,\\
\label{jbd6}
p={2+2\omega\over 4+3\omega}, \quad
q={2\over 4+3\omega}.
\end{gather}
The holographic bound for the particle horizon
is shown in Fig. \ref{phjordan}.
From Fig. \ref{phjordan}, we see the bound is violated at later time.
The holographic bound for the apparent horizon
is shown in Fig. \ref{ahjordan}.
From Fig. \ref{ahjordan}, we see that the bound is satisfied if it is satisfied 
initially.
In Einstein frame, the Brans-Dicke Lagrangian is 
\begin{equation}
\label{1}
{\cal L}= \sqrt{-g} \left[-\frac{1}{2\kappa^2} R
-\frac{1}{2}g^{\mu\nu}\partial_\mu \sigma \partial_\nu \sigma\right]
-{\cal L}_{m}(\psi, e^{-\alpha\sigma}g_{\mu\nu}).
\end{equation}
The above Lagrangian (\ref{1}) is from Eq. (\ref{bdlagr}) by the transformations
\begin{gather}
\label{conformala}
g_{\mu\nu}=e^{\alpha\sigma}\gamma_{\mu\nu},\\
\label{conformalb}
\phi={1\over G}e^{\alpha\sigma},
\end{gather}
where $\kappa^2=8\pi G$, $\alpha=\beta\kappa$, and $\beta^2=2/(2\omega+3)$. 
Remember that the Jordan-Brans-Dicke Lagrangian is not
invariant under the above transformations (\ref{conformala})
and (\ref{conformalb}).

The corresponding cosmological equations are
\begin{gather}
\label{3.1a}
H^2+{k \over a^2}={\kappa^2 \over 3}\left({\frac 1 2}\dot{\sigma}^2
+e^{-2\alpha\sigma}\rho\right),\\
\label{3.1b}
\ddot{\sigma}+3H\dot{\sigma}={\frac 1  2}\alpha e^{-2\alpha\sigma}\rho,\\
\label{3.1c}
\dot{\rho} +3H\rho={\frac 3 2}\alpha\dot{\sigma}\rho.
\end{gather}
Here we consider the solutions for the case $k=1$ and $p=0$ only.
With $8\pi G/3=1$, 
the initial conditions are
$$a(t_i)=\left[{\sqrt{2}(18+\alpha^2)\,t_i\over 4\sqrt{18-\alpha^2}}\right]^{
12/(18+\alpha^2)},\quad \sigma(t_i)={\alpha\over 3}\ln a(t_i).$$

\begin{figure}[htb]
\vspace{-0.1in}
\begin{center}
$\begin{array}{cc}
\epsfxsize=2.7in \epsffile{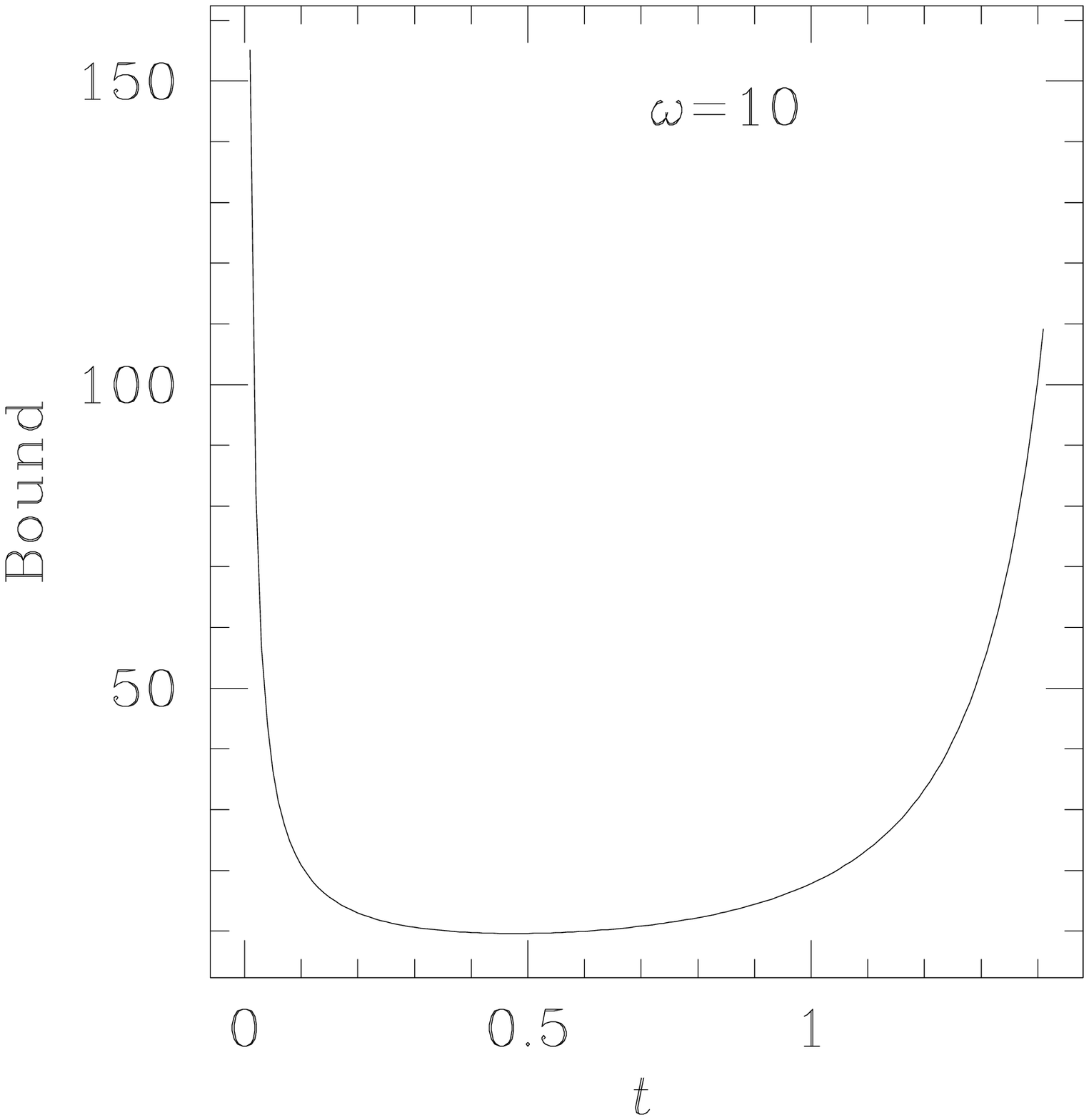}&
\epsfxsize=2.7in \epsffile{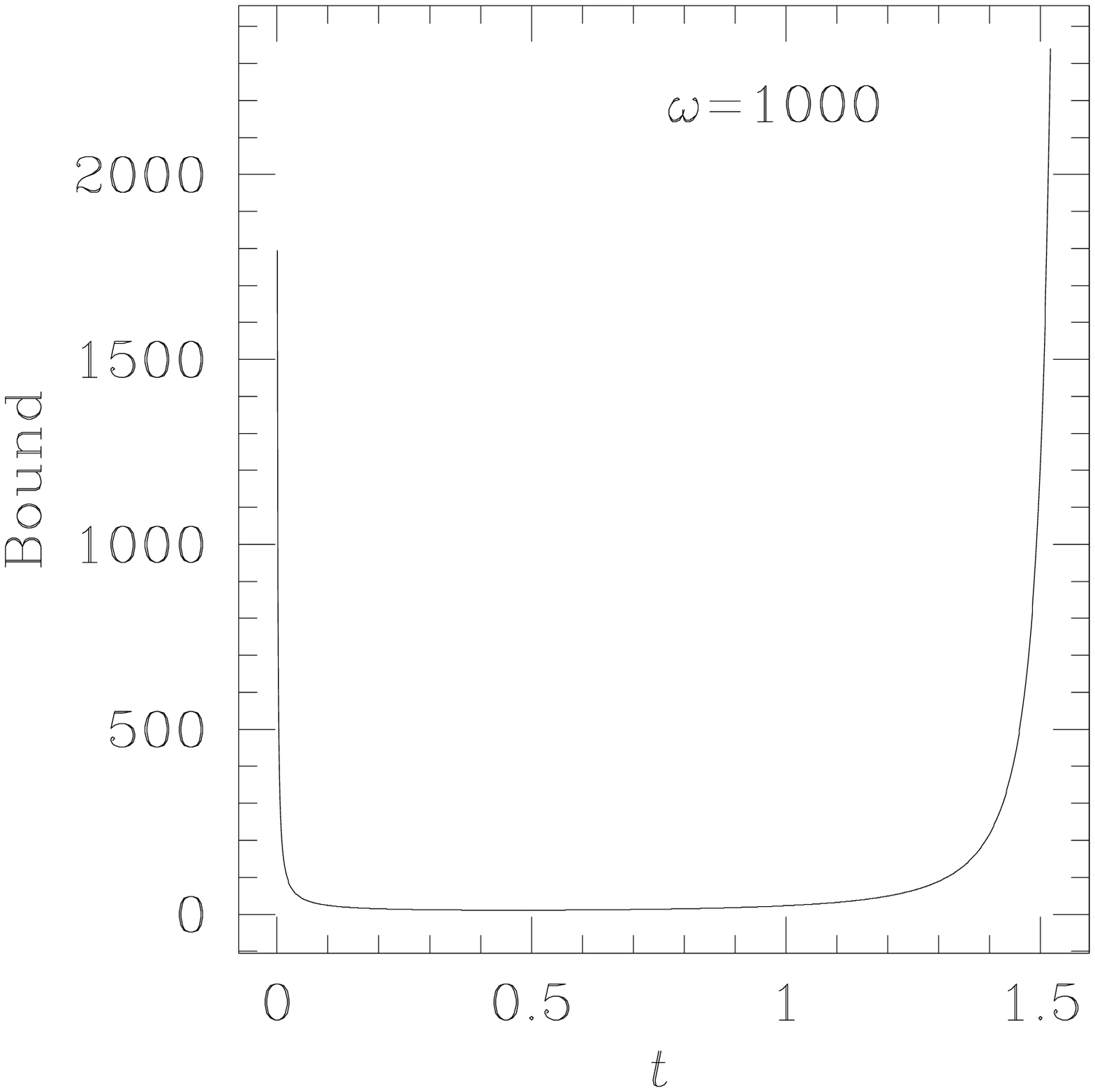}
\end{array}$
\end{center}
\vspace{-0.5in}
\caption{The holographic bound for particle horizon in Einstein frame
with $\omega=10$ and 
$\omega=1000$ respectively.}
\label{phpauli}
\end{figure}

\begin{figure}[htb]
\vspace{-0.1in}
\begin{center}
$\begin{array}{cc}
\epsfxsize=2.7in \epsffile{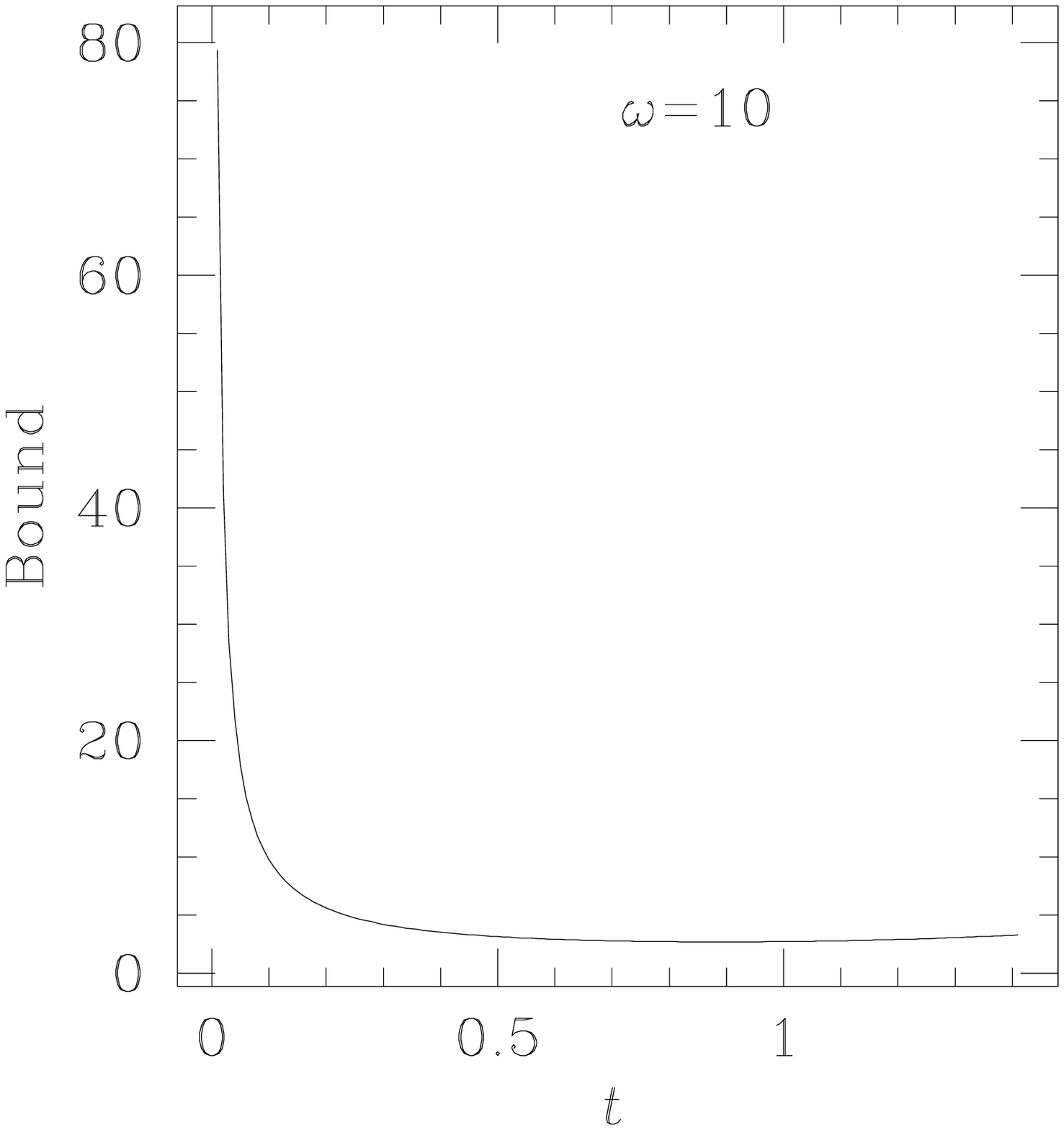}&
\epsfxsize=2.7in \epsffile{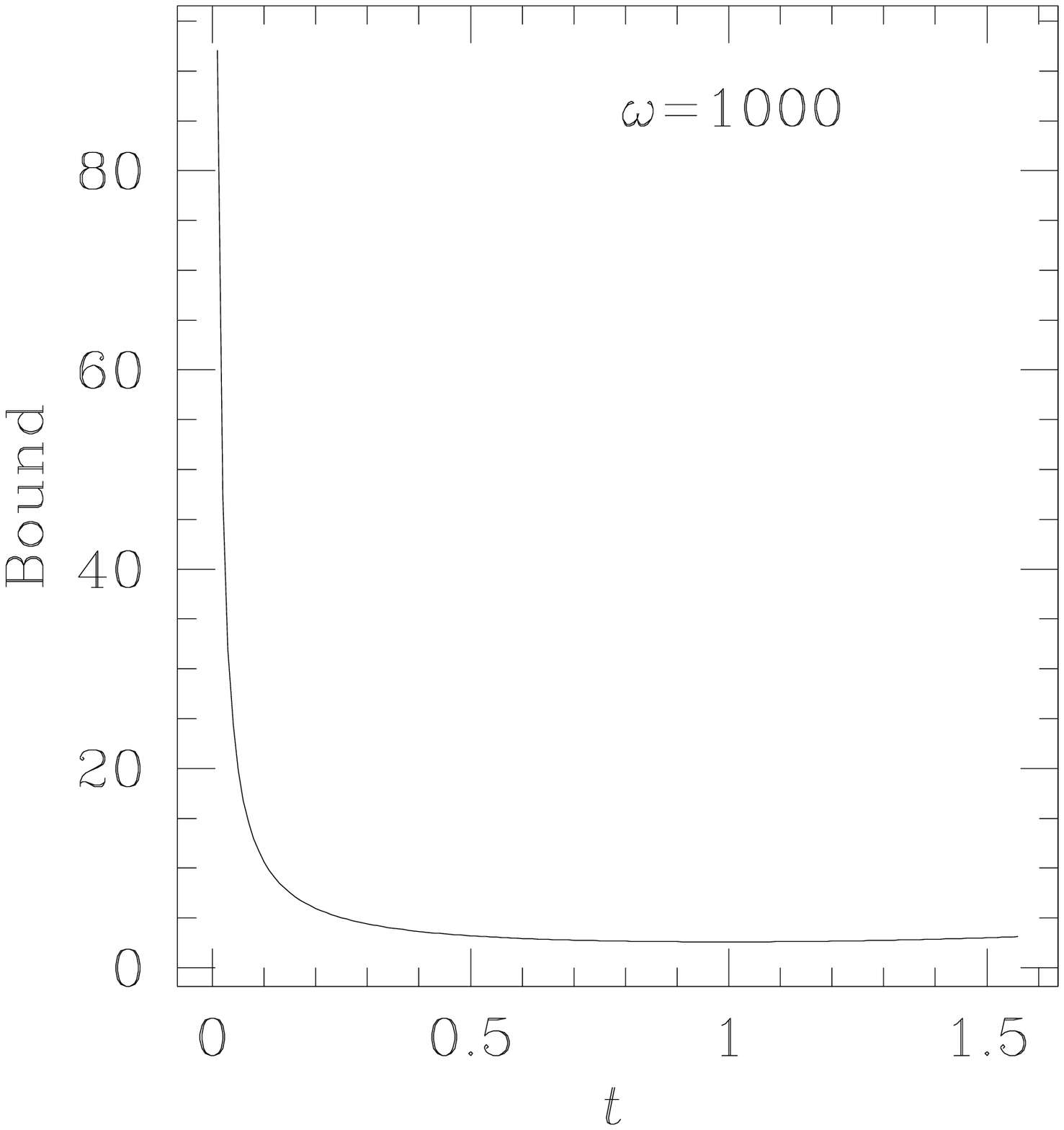}
\end{array}$
\end{center}
\vspace{-0.5in}
\caption{The holographic bound for apparent horizon in
Einstein frame with $\omega=10$ and 
$\omega=1000$ respectively.}
\label{ahpauli}
\end{figure}

The holographic bound for the particle horizon
is shown in Fig. \ref{phpauli}.
From Fig. \ref{phpauli}, we see that the bound is violated at later time.
The holographic bound for the apparent horizon is shown in Fig. \ref{ahpauli}.
From Fig. \ref{ahpauli}, we see that the bound is satisfied if it is satisfied 
initially. 

\section{Conclusions}

The holographic bound for Brans-Dicke cosmology 
is not satisfied if we use the particle horizon,
but it is satisfied if we use the apparent horizon.
we can understand the above result by using Bousso's conjecture:
In any spacetime satisfying Einstein's equation and 
the dominant energy condition, the total entropy $S$ contained
in any null hypersurface $L$ bounded by some connected ($D-2$) 
dimensional spacelike surface $B$ with area $A$ and generated
by null geodesics with non-positive expansion must satisfy
$S\le A/4G$. In Einstein frame, the spacetime satisfies Einstein's 
equation with $\rho_{tot}=e^{-2\alpha\sigma}\rho + \dot{\sigma}/2$
and $p_{tot}=e^{-2\alpha\sigma} p + \dot{\sigma}/2$. Therefore,
the matter source satisfies the dominant energy condition. Bousso's
covariant entropy conjecture tells us that the holographic bound is
satisfied in the region within the apparent horizon. In this paper,
we found that the holographic bound is satisfied for the $k=1$ matter
dominated Brans-Dicke universe. This can be taken as an evidence
to support Bousso's conjecture.

\bigskip
{\noindent \large \bf Acknowledgement}

\smallskip
The author would like to thank Raphael Bousso for
his helpful correspondence with the subject of Bousso's conjecture.


\begin{thebibliography}{gong}
\bibitem{gong} Y. Gong, gr-qc/9909013, to be appear in Phys. Rev. D, (2000).
\bibitem{fs} W. Fischler and L. Susskind, hep-th/9806039.
\bibitem{dbsr} D. Bak and S. Rey, hep-th/9902173.
\bibitem{nkal} N. Kaloper and A. Linde, hep-th/9904120, Phys. Rev. 
D {\bf 60}, 103509 (1999).
\bibitem{bousso} R. Bousso, hep-th/9905177, JHEP {\bf 9907}, 004 (1999); 
R. Bousso, hep-th/9906022, JHEP {\bf 9906}, 028 (1999).
\bibitem{fmw} E.E. Flanagan, D. Marolf and R.M. Wald, hep-th/9908070, see
also the talk given by R. Wald during this conference.
\bibitem{cosmo} S.K. Rama, Phys. Lett. B {\bf 457}, 268 (1999);
S.K. Rama and T. Sarkar, {\it ibid}, {\bf 450}, 55 (1999);
R. Easther and D. Lowe, Phys. Rev. Lett. {\bf 82}, 4967 (1999).
\end{thebibliography}
\end{document}